\documentclass[%
 reprint,
 amsmath,amssymb,
 aps,
floatfix,
]{revtex4-2}

\usepackage{graphicx}% Include figure files
\usepackage{dcolumn}% Align table columns on decimal point
\usepackage{bm}% bold math
\usepackage{cleveref}
\usepackage{tikz}
\usetikzlibrary{matrix}
\usepackage[normalem]{ulem}

\begin{document}

\preprint{APS/123-QED}

\title{Hermitian Nonlinear Wave Mixing Controlled by a PT-Symmetric Phase Transition}

\author{Noah Flemens}
\email{nrf33@cornell.edu}
\author{Jeffrey Moses}
\email{moses@cornell.edu}
\affiliation{School of Applied and Engineering Physics, Cornell University, Ithaca, New York 14853, USA}%

\date{\today}% It is always \today, today,

\begin{abstract}

While non-Hermitian systems are normally constructed through incoherent coupling to a larger environment, recent works have shown that under certain conditions coherent couplings can be used to similar effect. We show that this new paradigm enables the behavior associated with the PT-symmetric phase of a non-Hermitian subsystem to control the containing Hermitian system through the coherent couplings. This is achieved in parametric nonlinear wave mixing where simultaneous second harmonic generation replaces the role of loss to induce non-Hermitian behavior that persists through a full exchange of power within the Hermitian system. These findings suggest a new approach for the engineering of dynamics where energy recovery and sustainability are of importance that could be of significance for photonics and laser science.
\end{abstract}

\maketitle

Over the past two decades, the unique physics that emerge from open non-Hermitian systems have enabled a multitude of new device capabilities that overcome the limitations of their closed Hermitian equivalents \cite{El-Ganainy2018, Ozdemir2019, El-Ganainy2019Mar, Parto:21}. These capabilities arise largely through the dynamics that emerge near exceptional points in the non-Hermitian system eigenspectra, at which both eigenvalues and eigenvectors coalesce and regions of broken and unbroken PT-symmetry are demarcated under conditions of balanced gain and loss. Many device functionalities including single-mode lasing \cite{Hodaei2014Nov}, unidirectional invisibility \cite{Lin2011May, Feng2013Feb}, asymmetric mode-switching \cite{Ghosh2016Apr, Doppler2016Sep, Khurgin:21}, exceptional point enhanced sensitivity \cite{Hodaei2017Aug, Chen2017Aug, Lai2019Dec}, and improved efficiency and bandwidth of parametric amplification \cite{Ma:15, El-Ganainy:15, Zhong:16, Ma:17} have been proposed or realized through careful engineering of gain and loss.

However, the need for incoherent gain and loss creates practical limitations in non-Hermitian devices. It limits efficiency, creates inflexibility in the gain and loss bands, and produces undesirable signal-to-noise characteristics near exceptional points \cite{Lau2018Nov, Langbein2018Aug, Chen2019Aug}. To circumvent these limitations, recent works have investigated coherent interactions that can be used to the same effect. For instance, nonlinear parametric wave-mixing processes -- where there is coupling between modes of different frequency or polarization -- can be used to coherently add and remove energy from bosonic subsystems, thereby inducing non-Hermitian behavior without an incoherent exchange of energy with the medium \cite{Miri:16, Jiang:19, Zhang:20, Bergman:21, Wang:19, Roy:21, Roy2021Feb}. In these works, strong laser fields act as a reservoir of photons that can be exchanged with a subsystem. While these interactions usually take place in an approximately linear regime, in which the reservoir is effectively unperturbed, they have recently been extended to the nonlinear gain saturation regime where appreciable energy is added or removed from the strong driving fields \cite{Roy:21, Roy2021Feb}.

\begin{figure}[htbp]
	\centering
	   \includegraphics[width=3.4in]{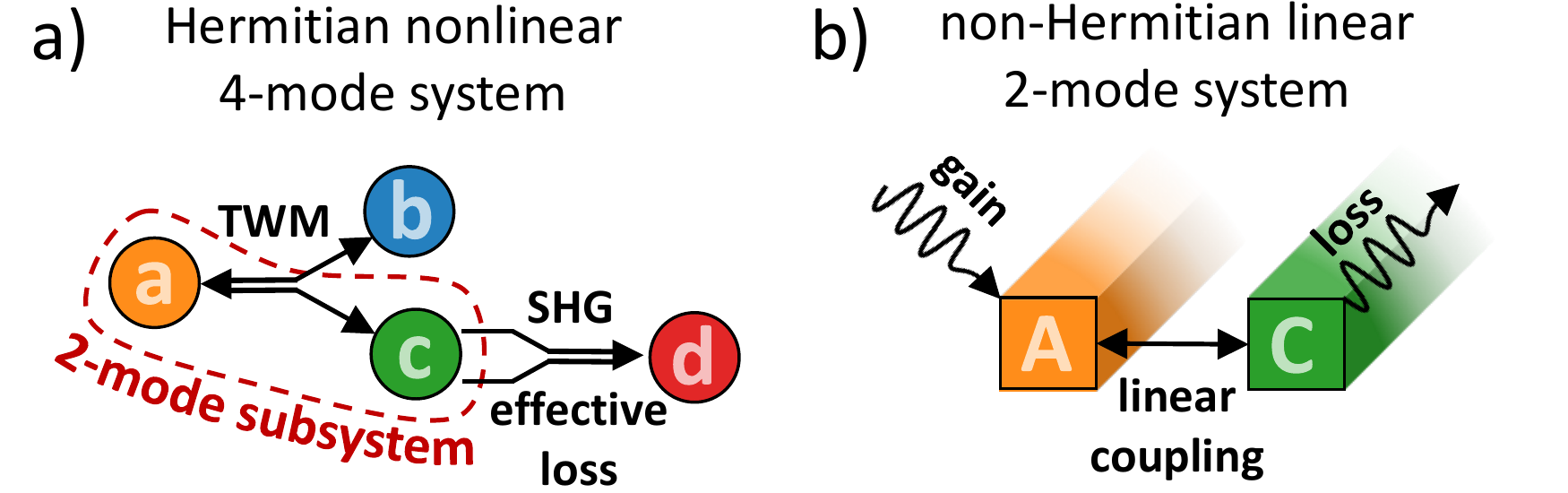}
        \caption{(a) A Hermitian system consisting of hybridized TWM and SHG can be represented by a 2-mode subsystem with an effective loss channel. This subsystem behaves analogously to (b) a non-Hermitian linear 2-mode system (e.g., coupled waveguides) with balanced gain and loss.}
    \label{fig:NHWaveguidesAnalogy}
\end{figure}

\begin{figure*}[htbp]
	\centering
	   \includegraphics[width=\textwidth]{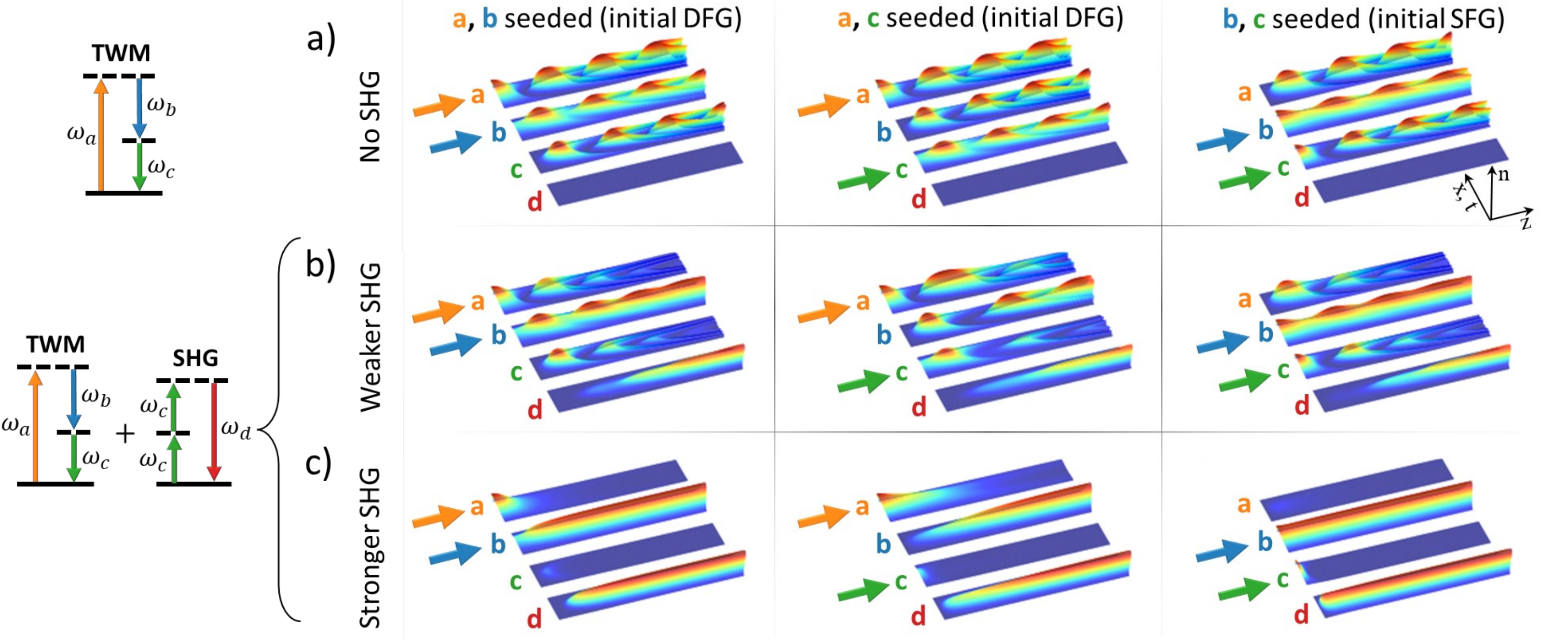}
        \caption{(left) Virtual energy level diagrams depicting (right) energy-conserving photon exchanges with only two of the fields $a$, $b$, and $c$ seeded (indicated by the arrows) for (a) phase-matched conventional TWM, where an oscillatory exchange of power occurs; and (b,c) when SHG ($2\omega_c = \omega_d$) is simultaneously phase matched. In (b), SHG is weak compared to TWM leading to a damped oscillatory conversion with asymptotic transfer to fields $b$ and $d$. In (c), SHG is strong compared to TWM, inhibiting SFG and leading to a unidirectional transfer of energy to fields $b$ and $d$. We note that the parameter $\eta_{\infty}$, as defined later, is held constant across columns in each row, implying differences in material constants and/or photon frequencies.}
    \label{fig:TWMDynamics}
\end{figure*}

Here we investigate the back-action of non-Hermitian subsystems on the behavior of the coherently coupled driving fields in such systems and find new phenomena. Our platform is simply parametric three-wave mixing (TWM) hybridized with second harmonic generation (SHG) (Fig. \ref{fig:NHWaveguidesAnalogy}a), where the SHG provides an effective loss channel on one of the fields.  In this Hermitian, 4-mode, nonlinear wave mixing system, we find a 2-mode subsystem that exhibits a PT-symmetric phase transition in analogy to coupled waveguides with balanced gain and loss (Fig. \ref{fig:NHWaveguidesAnalogy}b). Back-action on the enclosing system results in all modes evolving according to the subsystem PT-symmetric phase. This discovery offers exciting new avenues for extending applications of non-Hermitian physics to systems where high efficiency and energy conservation is desired -- a regime precluded for non-Hermitian devices due to their inherent lossy nature and/or requirement of an external gain source. In this case, breaking the cyclic nature of TWM allows for a unidirectional flow of photons that enables efficient frequency conversion. 

Conventional TWM interactions can be described as a cyclic exchange of photons between a higher frequency field at $\omega_a$ with two fields at lower frequencies $\omega_b$ and $\omega_c$ such that $\omega_a=\omega_b+\omega_c$. This one-to-two photon exchange is mediated by the quadratic nonlinear polarizability of a noncentrosymmetric medium and is energy-conserving when all frequencies are far from any material resonances \cite{Armstrong:62}. SHG is the degenerate case of TWM where the two lower frequencies are equal. In this paper, we consider the process $\omega_c + \omega_c = \omega_d$. For an efficient exchange of photons between fields, coherence between propagating and nonlinear polarization fields of the material must be maintained at each frequency. This occurs when the wave-vector mismatch, $\Delta \vec{k}_{abc} = \vec{k}_a-\vec{k}_b-\vec{k}_c$ for TWM and $\Delta \vec{k}_{dcc} = \vec{k}_d-2\vec{k}_c$ for SHG, vanishes (known as perfect phase matching) \cite{Giordmaine:62}.

For monochromatic plane waves, this hybrid system of TWM and simultaneous SHG can be modeled by four coupled evolution equations derived from Maxwell's equations. Hybridization is made possible by perfect phase matching of both processes (which can be achieved under various conditions \cite{Flemens:21}), resulting in:
\begin{subequations} \label{eq:system}
\begin{align}
    d_{z}u_a(z) &= i \Gamma_{abc} u_b(z) u_c(z)\label{eq:a} \\
    d_{z}u_b(z) &= i \Gamma_{abc} u_a(z) u_c^*(z)\label{eq:b} \\
    d_{z}u_c(z) &= i \Gamma_{abc} u_a(z) u_b^*(z) + 2i \Gamma_{dcc} u_d(z) u_c^*(z)\label{eq:c} \\
    d_{z}u_d(z) &= i \Gamma_{dcc} u_c^2(z). \label{eq:d}
\end{align}
\end{subequations}
\noindent The $u_j(z)$ are non-dimensional electric field amplitudes for $j \in \{a,b,c,d\}$ where $\left| u_j(z) \right|^2 = n_j(z)$ is the photon flux density in the $j$th field normalized by the total initial photon flux density, which we refer to as the fractional photon flux density of the $j$th field. $\Gamma_{abc}$ and $\Gamma_{dcc}$ are the drive strengths of the TWM and SHG processes, respectively. (Definitions in terms of complex electric field amplitudes $A_j(z)$, refractive indices, $\textrm{n}_j$, and nonlinear coefficient, $d_{\text{eff}}$: $k_j = \omega_j \textrm{n}_j / c$, $u_j(z)=\sqrt{2 \textrm{n}_j \epsilon_0 c/\hbar \omega_j F_0}A_j(z)$ where $F_0=\sum_j 2 \textrm{n}_j \epsilon_0 c\ |A_j(z=0)|^2/\hbar \omega_j$ is the total initial photon flux. $\Gamma_{ijk}=1/p\sqrt{\hbar \omega_i \omega_j \omega_k d_{ijk}^2 F_0/2 \textrm{n}_i \textrm{n}_j \textrm{n}_k \epsilon_0 c^3}$ where $p$ relates to the degeneracy of the process ($p=1$ for TWM and $p=2$ for SHG). $d_{ijk}$ is proportional to the tensor element of the quadratic electric susceptibility for the specific field polarizations of the three mixing fields $i$, $j$, and $k$.

Conventional TWM of waves with Gaussian transverse (spatial or temporal) mode profiles is depicted in Fig. \ref{fig:TWMDynamics}a. Conventional TWM takes place when $|\Delta k_{abc}| = 0$ and $|\Delta k_{dcc}| >> 0$, and in this case, Eqs. \ref{eq:system}c,d reduce to $d_{z}u_c = i \Gamma_{abc} u_a u_b^*$ and $d_{z}u_d = 0$. For any combination of two fields initially nonzero, we observe evolution that cycles between the processes of difference frequency generation (DFG) ($\omega_a \rightarrow \omega_b, \omega_c$), and sum frequency generation (SFG) ($\omega_b, \omega_c \rightarrow \omega_a$). Due to the nonlinear dependence on field amplitudes in Eqs. \ref{eq:system}, the periodicity of the conversion cycle varies across the transverse coordinate, leading to inhomogeneous conversion dynamics and a fundamental limitation on the conversion efficiency of the device as discussed in \cite{Flemens:21}.

When SHG is coupled to one of the lower frequency fields by satisfying $\Delta \vec{k}_{dcc} = 0$, SFG is inhibited and two distinct phases of dynamics are observed (Fig. \ref{fig:TWMDynamics}b,c). In both phases, we observe that asymptotically full conversion from modes $a$ and $c$ to $b$ and $d$ takes place independent of the initial local intensity, effectively homogenizing the modal transfer between the input and output fields. However, when TWM is strong in comparison to SHG (Fig. \ref{fig:TWMDynamics}b), the TWM is characterized by damped oscillations, and in contrast, when SHG is strong compared to TWM (Fig. \ref{fig:TWMDynamics}c), SFG is fully inhibited and the conversion is monotonic. In the following, we show analytically how the behavior of this closed Hermitian system and the emergence of these distinct phases results from underlying non-Hermitian physics. 

To begin this discussion, we point out that phase-matched SHG differs from most TWM interactions, in that the conversion dynamics are not cyclic. The displacement of photons to the second harmonic (SH) field is monotonic and irreversible, thus sharing a primary feature of loss due to contact with a thermal bath. This was pointed out in the context of parametric amplification \cite{Flemens:21}, in which SHG was observed to induce behavior normally associated with loss \cite{Ma:15, El-Ganainy:15, Zhong:16, Ma:17}. Yet, unlike a heat bath, the coupling between a wave and its SH is coherent, and unidirectional flow is a consequence of a vanishing polarization field at both the fundamental and SH frequencies. Moreover, since the growth of the SH field is quadratic in the fundamental (Eq. \ref{eq:d}), the irreversibility of flow is even insensitive to pi phase modulations in the fundamental field \cite{Flemens:21}. In the hybridized process in Fig. \ref{fig:TWMDynamics}b,c, this insensitivity enables a unidirectional flow of energy to the SH field even as the fundamental field undergoes conversion cycles. Thus, even when taken to full conversion, the SHG provides a loss channel for the TWM system by which we might expect the emergence of an exceptional point that demarcates regions of broken and unbroken PT-symmetric phases.

To begin the analysis, we derive from Eqs. \ref{eq:system} a set of equations that describe the rate at which photons are added and removed from each field by a given nonlinear interaction (TWM or SHG). These are derived by computing derivatives of each $n_j$:
\begin{subequations} \label{eq:rate}
\begin{align}
    d_{z}n_a(z) &= -\rho_{abc}(z) \label{eq:dna} \\
    d_{z}n_b(z) &= \rho_{abc}(z) \label{eq:dnb} \\
    d_{z}n_c(z) &= \rho_{abc}(z) - 2 \rho_{dcc}(z) \label{eq:dnc} \\
    d_{z}n_d(z) &= \rho_{dcc}(z) \label{eq:dnd}
\end{align}
\end{subequations}
\noindent where $\rho_{abc}(z) = 2 \Gamma_{abc} Im\{u_a^*(z) u_b(z) u_c(z)\}$ and $\rho_{dcc}(z) = 2 \Gamma_{dcc} Im\{u_d(z) (u_c^*(z))^2 \}$ represent the rate at which photons are added (removed) from each field by TWM and SHG, respectively. We have chosen the sign convention such that $\rho_{abc}(z)>0$ represents DFG and $\rho_{abc}(z)<0$ represents SFG. It is always the case that $\rho_{dcc}(z)\geq0$ since photons can only move unidirectionally into field $d$ from field $c$ as discussed above. From these equations it is simple to derive a set of linearly independent relationships known as the Manley-Rowe equations that define conserved quantities in terms of photon flux:
\begin{subequations} \label{eq:mr}
\begin{align}
    N_1 &= n_{a,0} + n_{b,0} = n_a(z) + n_b(z) \label{eq:mrab} \\
    N_2 &= n_{a,0} + n_{c,0} + 2 n_{d,0} = n_a(z) + n_c(z) + 2 n_d(z), \label{eq:mracd}
\end{align}
\end{subequations}
\noindent where the $n_{j,0}$ are the initial fractional photon flux densities and $n_{d,0}=0$ for the system under consideration. Thus, as the fields evolve, the number of photons in the $a$-$b$ and $a$-$c$-$d$ subsystems are constrained by the initial fractional photon flux density of the seeded fields. From this constraint and the unidirectionality of SHG, we can infer $n_b(z\rightarrow\infty)=N_1$ and $n_d(z\rightarrow\infty)=N_2/2$ while $n_a(z\rightarrow\infty)=n_c(z\rightarrow\infty)=0$ which captures the asymptotic behavior seen in Fig. \ref{fig:TWMDynamics}b,c.

We now seek to understand the intermediate dynamics seen in Fig. \ref{fig:TWMDynamics}b,c in terms of non-Hermitian physics. Since Eq. \ref{eq:mracd} suggests SHG acts directly to remove photons from fields $a$ and $c$, we direct our attention to the $a$-$c$ subsystem before turning our attention to the full system. Typically, an investigation of non-Hermitian physics involves computation of the eigenspectra for a linearly coupled subsystem with gain and loss. While nonlinear TWM systems have long been investigated in approximately linear regimes by way of undepleted field approximations or adiabatic elimination, here we take a new approach that allows us to investigate the non-Hermitian features in the fully nonlinear regime without approximation. This analysis requires \textit{a priori} knowledge of the evolution of fields $b$ and $d$ by first solving the wave mixing equations (Eqs. \ref{eq:system}) numerically. We can intuitively think of field $b$ as contributing to a propagation varying coupling constant $\kappa_{ac}(z) = \Gamma_{abc} u_b(z)$ for fields $a$ and $c$ while $\gamma_{cc}(z) = \Gamma_{dcc} |u_{d}(z)|$ represents a monotonically growing two-photon loss on field $c$. We then cast the $a$-$c$ subsystem (Eqs. \ref{eq:a},c) in a frame where the loss on mode $c$ is balanced by equivalent gain on mode $a$ by performing the gauge transformation $[u'_c, u'_a] = [u_c, u_a] e^{\int_0^{z}\gamma_{cc}(z')dz'}$. Over length scales where $|\Delta k_{abc}z|$ and $|\Delta k_{dcc}z| \ll \pi$, we find this transformation provides a powerful analytic tool for identification of parameters that dictate the occurrence of phase transitions within the nonlinear system.

Substituting these coordinates into Eqs. \ref{eq:a} and \ref{eq:c}, we can write the coupled $a$-$c$ subsystem equations in the simplified Hamiltonian form,
\begin{gather}
    -i\frac{d}{dz}
    \begin{bmatrix}
        u'_c(z) \\
        u'_a(z)
    \end{bmatrix}
    =
    \begin{bmatrix}
        i\gamma_{cc}(z)& \kappa_{ac}^*(z) \\
        \kappa_{ac}(z) & -i\gamma_{cc}(z) 
    \end{bmatrix}
    \begin{bmatrix}
        u'_c(z) \\
        u'_a(z)
    \end{bmatrix}.
    \label{eq:two-level}
\end{gather}
\noindent The propagation-dependent Hamiltonian of this system is given by 
%$H_{ac}(z)=\kappa(z)\sigma_x - i\gamma(z)\sigma_z$
$H_{ac}(z) = \vec{g}(z) \cdot \vec{\sigma}$ where $\vec{g}(z) = (Re \{\kappa_{ac}(z)\}, Im\{\kappa_{ac}(z)\}, i \gamma_{cc}(z))$ expresses the coupling and loss of the system and $\vec{\sigma}$ is the Pauli vector. This Hamiltonian is non-Hermitian except in the $\gamma_{cc}(z)=0$ case, which represents conventional TWM without SHG. It is also easy to check $H_{ac}(z)$ commutes with the parity-time operator by computing $[H_{ac}(z),PT] = 0$ with parity inversion of fields $a$ and $c$ given by $P=\sigma_x$ and time reversal given by complex conjugation ($TuT^{-1}=u^*$).

\begin{figure}[htbp]
	\centering
	   \includegraphics[width=3.46in]{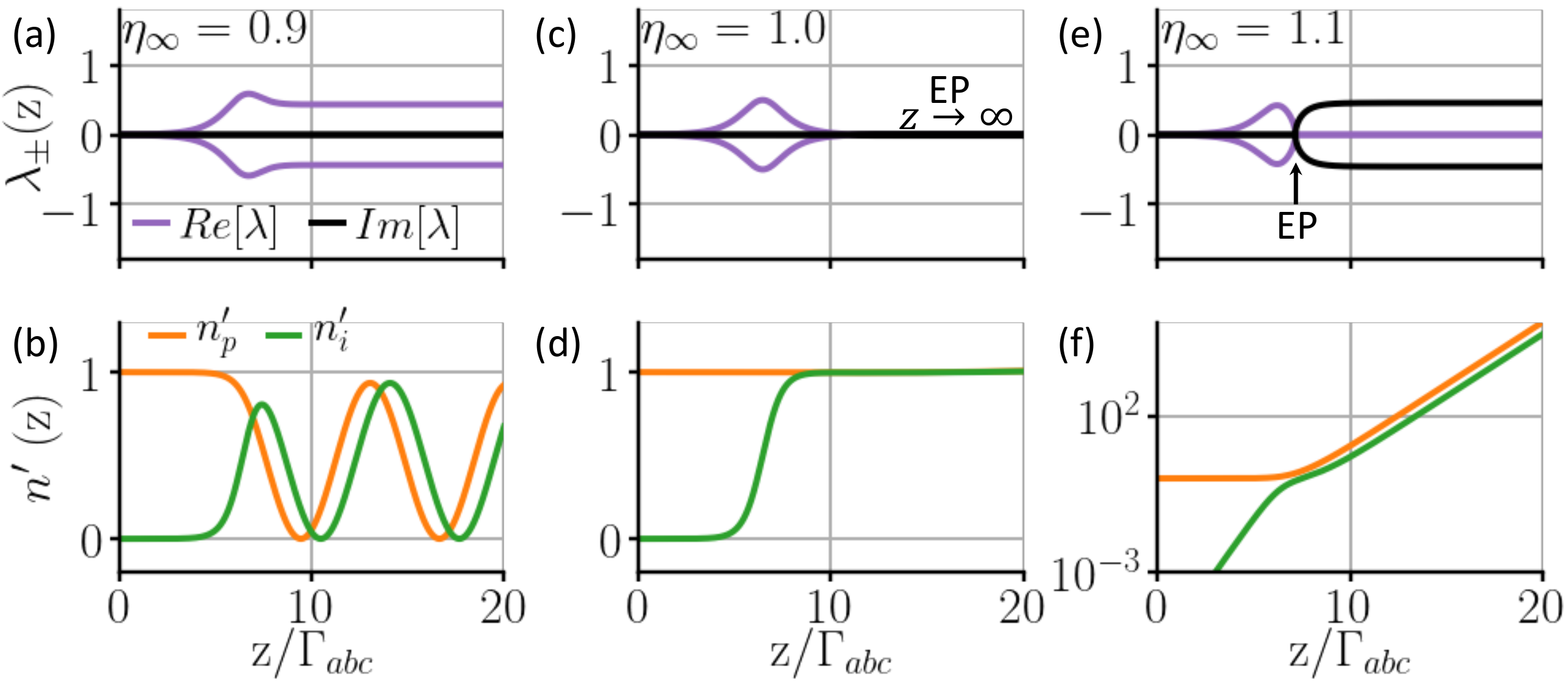}
        \caption{Numerical solutions for $a$-$c$ subsystem dynamics in the gauge transformed frame for (a,b), $\eta_{\infty}<1$, having power oscillations and purely real eigenvalues for all $z$; (c,d), $\eta_{\infty}=1$, in which fields $a$ and $c$ coalesce as the system asymptotically approaches the exceptional point; and (e,f), $\eta_{\infty}>1$, showing exponential growth and a transition from purely real to purely imaginary eigenvalues at the exceptional point. All cases: $n_{b,0}=10^{-5}$ and $n_{a,0}=1-n_{b,0}$.
        }
    \label{fig:PTSymmetry}
\end{figure}

A local eigenspectra analysis of $H_{ac}(z)$ yields propagation dependent eigenvalues $\lambda_\pm(z) = \pm \sqrt{\vec{g}(z) \cdot \vec{g}(z)}= \pm \sqrt{|\kappa_{ac}(z)|^2 - |\gamma_{cc}(z)|^2}$ where an exceptional point occurs when the coupling and loss are equivalent, or in physical terms, when the SHG and TWM processes act with equal strength on the subsystem. We introduce the state parameter $\eta(z) = \sqrt{|\gamma_{cc}(z)|/|\kappa_{ac}(z)|}$ which quantifies the relative strength of loss due to SHG to coupling due to TWM. We find the local right eigenvectors depend solely on this new parameter: $\vec{v}_\pm(z) = 1/\sqrt{2}[1, i \eta(z) \pm \sqrt{1-\eta(z)^2}]^T$. An exceptional point exists at $\eta=1$, where the eigenvalues and local right eigenvectors coalesce. When $\eta(z) < 1$, coupling by TWM is the dominant process and the subsystem is PT-symmetric with purely real eigenvalues. When $\eta(z) > 1$, loss of photons by SHG is dominant and PT-symmetry is broken, resulting in purely imaginary eigenvalues. Whether the system converges to the broken or unbroken PT-symmetric phase is determined by the steady state parameter $\eta_{\infty} \equiv \eta(z \rightarrow \infty)=\Gamma_{dcc} / \Gamma_{abc} \sqrt{N_2/2N_1}$. For values of $\eta_{\infty} <1$, fields $a$ and $c$ will forever engage in bidirectional power exchange via SFG/DFG conversion cycles (Fig. \ref{fig:PTSymmetry}a, b). When $\eta_{\infty}>1$, PT-symmetry is broken at finite $z$, leading to a complete elimination of the bidirectional exchange of power between fields $a$ and $c$ that is characteristic of TWM (Fig. \ref{fig:PTSymmetry}e, f). Thus, the conditions for PT-symmetry breaking are determined by the relative drive strength of the SHG and TWM processes and by the initial conditions through the conserved quantities of Eqs. \ref{eq:mr}.

\begin{figure*}[htbp]
	\centering
	   \includegraphics[width=\textwidth]{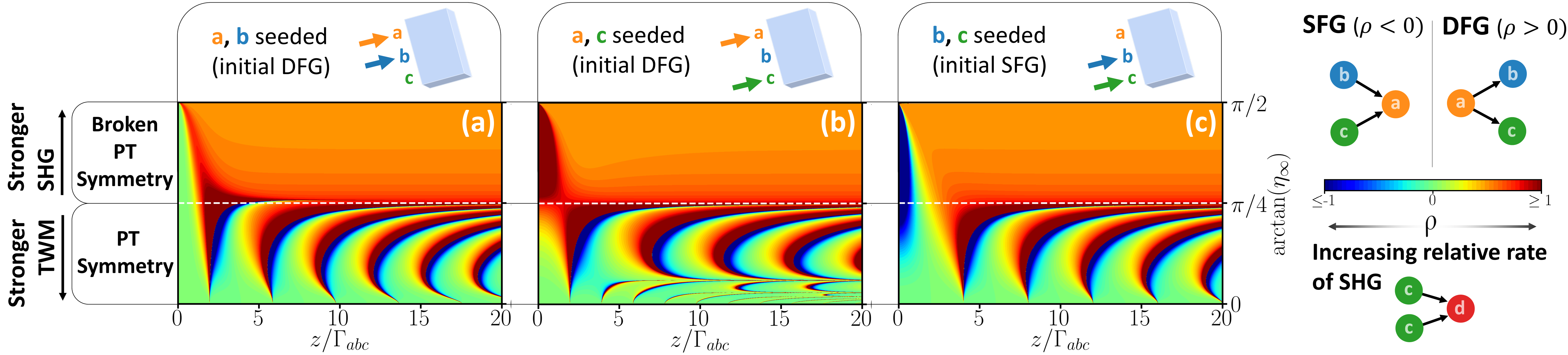}
        \caption{Gauge-invariant parameter $\rho(z)$ (color map) for initial conditions corresponding to DFG with (a) fields $a$ and $b$ seeded, (b) fields $a$ and $c$ seeded, and (c) SFG with fields $b$ and $c$ seeded. A white dashed line corresponding to $\eta_\infty = 1$ demarcates a sudden transition in behavior corresponding to the PT-symmetric phase of the $a$-$c$ subsystem. The colorscale is truncated at $\pm 1$ for clarity.}
    \label{fig:rho}
\end{figure*}

So far, we have revealed how non-Hermitian physics emerges in the $a$-$c$ subsystem of any TWM process as a consequence of simultaneously phase-matched SHG. Our approach also allows an exact analysis over the full range of power exchange dynamics, rather than employing a linearized model that excludes the coherence between the subsystem and external fields. Thus, we can analyze another interesting feature: the back-action of the non-Hermitian $a$-$c$ subsystem on the external fields $b$ and $d$. In the following analysis, we investigate how the abrupt transition in PT-symmetric phase in the non-Hermitian 2-mode $a$-$c$ subsystem imprints on the 4-mode Hermitian system, leading to the dynamics in Fig. \ref{fig:TWMDynamics}.

A compact representation of the 4-mode system dynamics is given by the real parameter $\rho(z) = \rho_{dcc}(z)/\rho_{abc}(z) = \eta \sqrt{n'_c/n'_a}$. Its magnitude quantifies the relative rate of photons being exchanged by SHG compared to TWM. At its extrema, $|\rho(z)|\in\{0, \infty\}$, only TWM or SHG occur, respectively. The sign of $\rho(z)$ represents the direction of TWM photon exchange, with $\rho(z)>0$ corresponding to DFG and $\rho(z)<0$ corresponding to SFG. Since this parameter is defined as a ratio, it is gauge invariant, and thus can be interpreted in terms of either the 2-mode $a$-$c$ subsystem or the full 4-mode system.

Figure \ref{fig:rho} depicts the photon exchange dynamics for three representative cases of TWM and in each case exhibits an abrupt phase transition in the dynamics of the closed Hermitian system. This phase transition is demarcated by the white dashed line at $\arctan(\eta_{\infty}) = \pi/4$ (i.e., $\eta_{\infty} = 1$). When $0 \leq \arctan(\eta_{\infty}) < \pi/4$, the $a$-$c$ subsystem always remains PT-symmetric and there is a perpetual oscillation in the relative rate of SHG and TWM interactions with the TWM process periodically switching between DFG and SFG, as seen in Fig. \ref{fig:TWMDynamics}b. In this phase, the dynamics of photon exchange are highly dependent on the specific system parameters. However, when $\pi/4 \leq \arctan(\eta_{\infty}) < \pi/2$, the exceptional point is crossed at finite $z$ and we see rapid convergence to dynamics involving only DFG and SHG at a fixed rate, leading to the monotonic growth of fields $b$ and $d$,  independent of which two of the TWM fields are initially seeded, as seen in Fig. \ref{fig:TWMDynamics}c. Thus, we see how the abrupt transition in the PT-symmetric phase of the non-Hermitian 2-mode subsystem is imprinted on the enclosing Hermitian 4-mode system. In fact, the exact relative rate of DFG and SHG reached in equilibrium,  $\rho(z \to \infty) = \eta_{\infty}^2 - \sqrt{\eta_{\infty}^4 - \eta_{\infty}^2}$, depends only on the $a$-$c$ subsystem state parameter $\eta_{\infty}$, and is independent of the initial behavior of the TWM system.

These findings have a number of interesting implications. We have shown that the evolution behavior of a Hermitian wave-mixing system is tied to non-Hermitian behavior of an enclosed subsystem, even through a full nonlinear power exchange. In non-Hermitian nonlinear wave mixing systems, it has been established that conversion between modes can be homogenized across their transverse profiles, linearizing the input-output behavior. This is important, as it solves the longstanding problem of inefficient frequency conversion \cite{Ma:15, El-Ganainy:15, Zhong:16, Ma:17, Flemens:21}. That this can be achieved without any real loss or coupling to a thermal bath -- enabled rather by coherent coupling to a co-propagating wave -- can circumvent other problems. As pointed out previously, thermal loading can be avoided \cite{Flemens:21} and phase noise improved \cite{Roy:21, Roy2021Feb}, and the loss band can be chosen by phase-matching technique rather than being tied to material or structural resonances. Moreover, in our system, the photons displaced from the enclosed subsystem are preserved in a coherent field that can be used in subsequent applications. All of these are significant capabilities for advancing frequency conversion technology that can be important for laser science and integrated photonics. More generally, the use of non-Hermitian physics to explicitly control an enclosing Hermitian system, as shown here, may have broader applicability and importance where energy efficiency and sustainability are of concern.

\begin{acknowledgments}
The authors would like to thank Francesco Monticone for useful discussions. This work was supported equally by the Cornell Center for Materials Research with funding from the NSF MRSEC program (DMR-1719875), which provided initial support, and by the NSF under grant no. ECCS-1944653. Underlying data are available at Ref. \cite{Repository}.
\end{acknowledgments}

%\bibliography{NHTWMbib} % Produces the bibliography via BibTeX.

\providecommand{\noopsort}[1]{}\providecommand{\singleletter}[1]{#1}%

\end{document}